\documentclass[useAMS]{mn2e}
\usepackage{rotating}
\usepackage{mncite}

\def\approxgt{\ifmmode \rlap{$>$}{}_{{}_{{}_{\textstyle\sim}}} \else%
$\rlap{$>$}{}_{{}_{{}_{\textstyle\sim}}}$\fi} 
\def\approxlt{\ifmmode \rlap{$<$}{}_{{}_{{}_{\textstyle\sim}}} \else%
$\rlap{$<$}{}_{{}_{{}_{\textstyle\sim}}}$\fi}

\LARGE \normalsize \title[Chandra observations of the neutron star
soft X--ray transient RX~J170930.2-263927]{Chandra observations of the
  neutron star soft X--ray transient RX~J170930.2-263927 returning to
  quiescence}

\author[P.G. Jonker et al.]
{P.G. Jonker$^1$\thanks{email : peterj@ast.cam.ac.uk},
M. M\'endez$^2$, 
G. Nelemans$^1$, 
R. Wijnands$^3$, 
M. van der Klis$^4$ \\
$^1$Institute of Astronomy, Madingley Road, CB3 0HA, Cambridge\\
$^2$SRON, National Institute for Space Research, Sorbonnelaan 2, 3584~CA Utrecht, 
The Netherlands\\ 
$^3$School of Physics and Astronomy, University of St Andrews, North Haugh, St Andrews, KY16 9SS, Fife, Scotland, UK\\
$^4$Astronomical Institute ``Anton Pannekoek'',
University of Amsterdam, Kruislaan 403, 1098 SJ Amsterdam\\ } 

\begin{document}

\maketitle

\begin{abstract}
\noindent 
We present our analysis of {\it Chandra} observations obtained when
the soft X--ray transient RX~J170930.2-263927 (XTE~J1709--267)
returned to quiescence after an outburst. Using the type I burst peak
luminosity found by Cocchi et al. (1998) and the value of N$_H$ we
derived from our spectral fits, the distance to RX~J170930.2-263927 can
be constrained to 2.5--10 kpc.  RX~J170930.2-263927 is probably
associated with the low--metalicity Globular Cluster NGC~6293, which
has a tidal radius of 14.2 arcminutes, since the projected distance to
the centre of the cluster is approximately 25 parsec (9--10
arcminutes). If the association is correct, RX~J170930.2-263927 would
be at $\sim$8.5 kpc. We determined that $\frac{{\rm
    L_{outburst}}}{{\rm L_{quiescence}}}\approxgt10^5$ for this
outburst. If the quiescent luminosity is caused by cooling of the
neutron star core then enhanced core cooling processes were at work
implying a neutron star mass of $\sim$1.7--1.8 M$_\odot$.  Combining
our {\it Chandra} observations with archival ROSAT observations we
show that the source most likely exhibits periods of sustained
low--level accretion.  Variable, low--level activity could provide an
alternative explanation for some of the sources in the recently
proposed category of {\it faint} soft X--ray transients.  We found
excess emission at $\sim$0.6 keV. If such an excess is a unique
feature of ultracompact systems, as was recently proposed,
RX~J170930.2-263927 must have a short orbital period as well. From the
constraints on the distance and the non--detection of the optical
counterpart with m$_V<20.5$, we conclude that this system must have a
low--mass companion.

\end{abstract}


\section{Introduction}
\label{intro}
Low--mass X--ray binaries (LMXBs) are binary systems in which a
$\approxlt$1\,$M_{\odot}$ star transfers matter to a neutron star or a
black hole. A large fraction of the LMXBs are transient, the so called
soft X--ray transients (SXTs; see \pcite{1997ApJ...491..312C}).
Characterising properties of SXTs are i) the accretion rate drops
several orders of magnitude when the source returns to quiescence
(\pcite{1984heta.conf...49V}) and ii) the spectra become soft when in
outburst. 

There are several mechanisms which can produce the low--level X--ray
emission observed in quiescence (see \pcite{1998A&ARv...8..279C} for
an overview). Firstly, mass accretion may be ongoing at a low level
possibly via an ADAF type flow (e.g. \pcite{1994ApJ...428L..13N}).
Secondly, provided the neutron star has a substantial magnetic field
(\approxgt$10^{7-8}$ Gauss) and a short spin period (milliseconds),
centrifugal forces may prevent accretion onto the neutron star (the
propeller mechanism; \pcite{1975A&A....39..185I}).  Shocks produced by
this propeller mechanism are then responsible for the X--ray emission.
Thirdly, the switch--on of a radio pulsar mechanism may produce
X--rays (\pcite{2000ApJ...541..849C}).  Finally, the neutron star core
and crust are heated by accretion episodes during outburst; subsequent
cooling of the neutron star in quiescence produces (soft) X--rays,
(e.g. \pcite{1987A&A...182...47V}; \pcite{1998ApJ...504L..95B}).

SXTs have been studied both in outburst and in quiescence with various
satellites (e.g. with ROSAT \pcite{1994A&A...285..903V}, with ASCA
\pcite{1998PASJ...50..611A}, with RXTE \pcite{2001ncxa.conf...94S},
with Beppo{\it SAX} \pcite{2001ESASP.459..463I}). From these studies a
basic picture of the outburst mechanism has emerged; the
thermal--viscous disk instability model (\pcite{1983ApJ...272..234S},
\pcite{1995ApJ...454..880C}, \pcite{1996ApJ...464L.139V}; for a recent
review and a more detailed account see \pcite{2001NewAR..45..449L}).

Recent observations of neutron star SXTs in quiescence with the {\it
  Chandra} and XMM--{\it Newton} satellites have fueled the debate on
the nature of the quiescent X--ray emission (e.g. see
\pcite{2002ApJ...573L..45W} and \pcite{2002ApJaccepted}). Furthermore,
using data obtained with the Beppo{\it SAX} satellite
\scite{1999ApL&C..38..297H} and \scite{2001ESASP.459..463I} suggested
that $\sim$10 bursting neutron stars form a separate class of {\it
  faint} SXTs.  Some may have a low peak--luminosity (typically
$\sim10^{36.5}$ erg s$^{-1}$), possibly like the accreting millisecond
X--ray pulsar SAX~J1808.4--3658 (\pcite{1998Natur.394..344W};
\pcite{1998Natur.394..346C}).  Later, \scite{2000MNRAS.315L..33K} showed 
that a class of faint SXT is expected on evolutionary grounds. These are
systems which have evolved beyond the period minimum of $\sim$80
minutes to orbital periods of 80--120 minutes.

Recently, \scite{2001ApJ...560L..59J} proposed that the excess
emission near 0.6 keV, unaccounted for by continuum models, found in a
handful of X--ray binaries (among which 4U~1626--67, which has an
orbital period of $\sim$41 minutes; \pcite{1995ApJ...449L..41A}) can
be explained by varying the absorption abundances of circumstellar O
and Ne with respect to solar. The abundances and the ultra--short
orbital periods of several of these systems can be explained providing
these systems are accreting from a degenerate companion
(\pcite{2001ApJ...560L..59J}; see \pcite{2002A&A...388..546Y} for
detailed evolutionary considerations).  Recently, two other transient
accreting millisecond X--ray pulsars have been found, XTE~J1751--305
(\pcite{2002IAUC.7867....1M}; \pcite{2002ApJ...575L..21M}) and
XTE~J0929--314 (\pcite{2002IAUC.7900....2G};
\pcite{2002ApJ...576L.137G}) with orbital periods of $\sim$42 and
$\sim$43 minutes, respectively. These systems fit--in with the class
of ultra--short period binaries. However, despite the fact that these
millisecond pulsars have short orbital periods XMM--{\it Newton}
observations of XTE~J1751--305 and {\it Chandra} observations with the
Low Energy Transmission Grating Spectrometer of XTE~J0929--314 did not
show evidence for excess emission near 0.6 keV, complicating the
picture (\pcite{2002ApJMiller1751}; \pcite{2002ApJsubm}).

RX~J170930.2--263927 was discovered by the ROSAT All Sky Survey
(\pcite{1999A&A...349..389V}). In 1997 the source was found in
outburst with the Proportional Counter Array onboard the RXTE
satellite (\pcite{1997IAUC.6543....2M}). Using Beppo{\it SAX}'s Wide
Field Camera, \scite{1998ApJ...508L.163C} found that the source
exhibits bursts, probably type~I X--ray bursts, which would establish
the compact object in RX~J170930.2--263927 as a neutron star.

In this paper we present results from {\it Chandra} observations of
the SXT RX~J170930.2-263927 obtained during the decay after an X--ray
outburst.

\section{Observations and analysis}
\label{analysis}
We observed RX~J170930.2--263927 five times with the Advanced CCD
Imaging Spectrometer (ACIS) back--illuminated S3 chip onboard the {\it
  Chandra} satellite (\pcite{1988SSRv...47...47W}). A log of the
observations is given in Table~\ref{log}. To mitigate the effects of
pile--up the ACIS S3 chip was read out in {\it continuous clocking}
mode during the first four observations (CC--mode, providing a time
resolution as high as 2.85 ms) and in a windowed timed exposure mode
during the last observation (TE--mode; time resolution of 2.24 s).

\begin{table*}
\caption{Log of the observations.}
\label{log}
\begin{center}

\begin{tabular}{lccc}
\hline
Observation & Observation date  & MJD  & Total effective \\ 
\# / ID  &  and start time (TT)  & (UTC) &  on source time (ksec.) \\
\hline
\hline
1 / 3462 & 12--03--2002 13:34 & 52345.565 & $\sim$14.2 \\
2 / 3463 & 18--03--2002 11:25 & 52351.475 & $\sim$5.6 \\
3 / 3464 & 01--04--2002 00:27 & 52365.018 & $\sim$5.1 \\
4 / 3475 & 10--04--2002 17:28 & 52374.727 & $\sim$6.0 \\
5 / 3492 & 25--04--2002 21:35 & 52387.898 & $\sim$5.2 \\

\end{tabular}
\end{center}

\end{table*}

The data were processed by the {\it Chandra} X--ray Center; events
with ASCA grades of 1, 5, and 7 were rejected. We used the standard
{\it CIAO} software to reduce the data (version 2.2.1). During the
second observation a background flare occured after $\sim$5.6 ksec
which lasted for the rest of the observation.  Therefore, we only used
the first $\sim$5.6 ksec of the total $\sim$15 ksec of this
observation in our analysis. In the first two observations dips in the
source count rate at periods of 1000 seconds and $\sim$700 seconds
were found. These dips are caused by the spacescraft dither, moving
the source over a bad pixel or a CCD node boundary. Events in such a
dip have been discarded. At the end of the first observation an X--ray
burst occured. We fitted the burst data separately (see
Section~\ref{results}). In order to study the timing properties of the
source the recorded event times were corrected to the approximate
photon arrival times using the method explained by
\scite{2001ApJ...563L..45P}.

The spectrum of the background in TE--mode ACIS--S observations
displays an emission feature just below 2 keV (see figure 6.15 of the
{\it Chandra} Proposers' Observatory Guide v.4 \footnote{available at
  http://asc.harvard.edu/udocs/docs/docs.html}). Furthermore, the
response of the high resolution mirrors jumps at $\sim$2 keV (due to
an iridium edge) and in CC--mode the background levels are 1024 times
higher than in TE--mode making the background emission feature just
below 2 keV more significant. All these effects make it very difficult
to distinguish source and background features and to model the
background and effective area of the telescope in the spectral area
near 2 keV.  Therefore, we excluded the 1.75--2.15 keV energy range
from our spectral analysis.  The spectra were extracted with 10 counts
per bin except for the spectrum in the last exposure which has only 5
counts per bin. The CC--mode spectral response is not calibrated but
in principle there should be no major differences with respect to the
TE--mode spectral response (this has been verified by
\pcite{2001ApJ...563L..45P}). We excluded energies below 0.35 and
above 8 keV from our spectral analysis since the TE--mode spectral
response is not well calibrated for those energies.

\section{Results}

\label{results}
\subsection{Source position and optical observations}

Observation 5 was obtained using the TE--mode. This provided us with
an X--ray image of the region around RX~J170930.2--263927. We detected
only one source at a count rate of (2.4$\pm$0.2)$\times10^{-2}$ counts
per second. We corrected the aspect solution using the {\it CIAO} tool
{\sc fix offsets}; next using {\sc celldetect} we obtained an accurate
position for RX~J170930.2--263927 of R.A.  $= 17h09m30.4s$, Decl
$=-26^\circ39'19.9''$ (uncertainty 0.6$''$, equinox 2000.0).

\begin{figure*}
  \includegraphics[width=15cm]{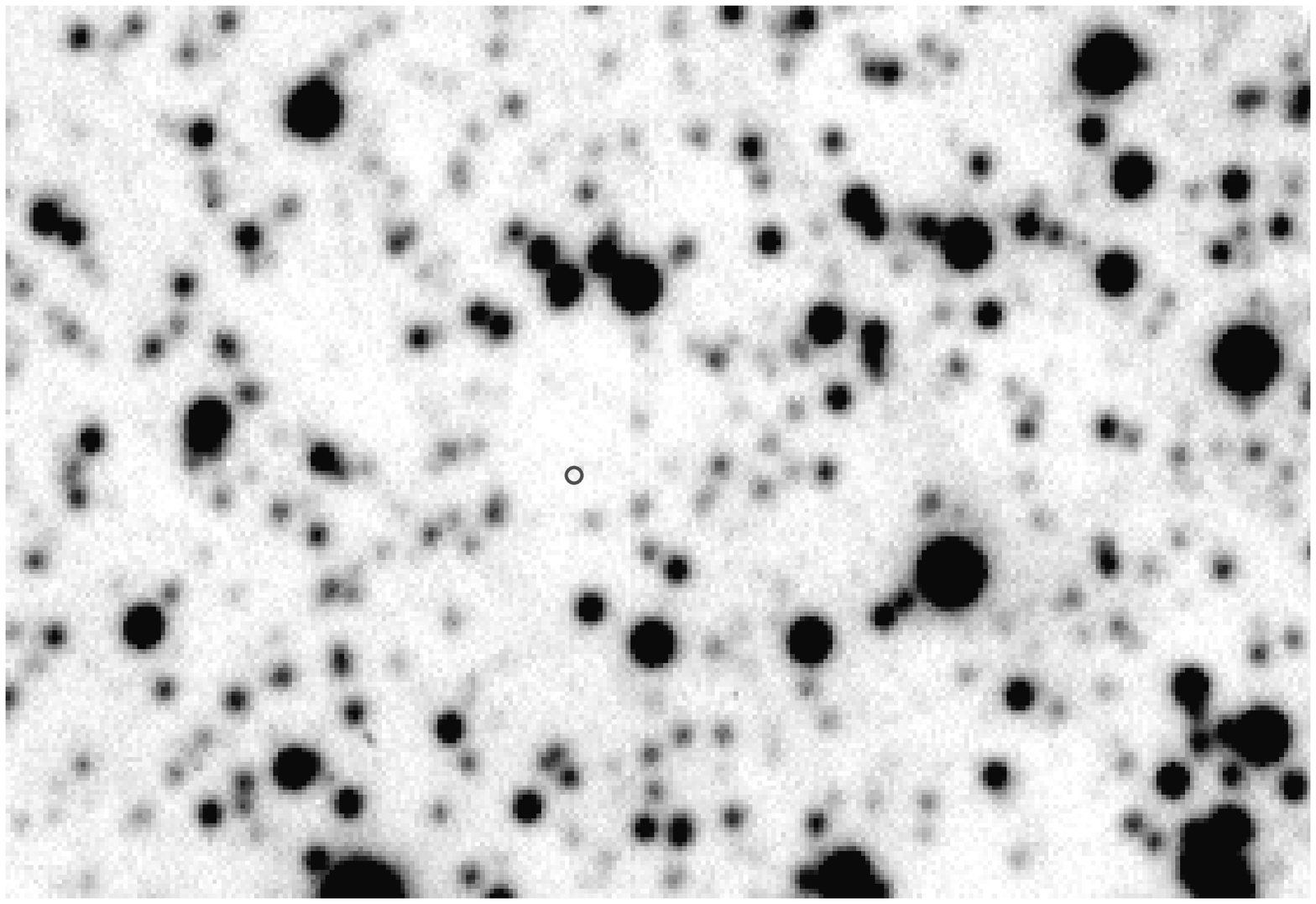}
\caption{V--band image ($\sim100''\times60''$; 5 minute
  integration) of RX~J170930.2--263927 obtained with the 2.5~m INT on
  La Palma.  North is up and East is to the left. The {\it Chandra}
  error circle is overplotted.}
\label{finder}
\end{figure*}

Given the accurate position and the relatively low N$_H$ towards the
source we obtained a 5 minute V--band image with the 2.5~m INT
telescope on La Palma on June 6, 2002 to search for the optical
counterpart of RX~J170930.2--263927. We note that by that time the
X--ray source was most likely back in quiescence. An astrometrical
solution for the optical image was obtained using the USNO positions
of 5 stars. The rms uncertainty of the astrometrical solution was
0.13$''$.  Hence, the overall uncertainty in the astrometry is
dominated by the {\it Chandra} positional uncertainty.  In
Figure~\ref{finder} we show the central part of the V--band image with
the Chandra error--circle overplotted.  Comparing the optical V--band
image with an older DSS R--band image we find no new sources with a
5~$\sigma$ upper limit magnitude of 20.5, nor did we find a strongly
variable source in or near the {\it Chandra} error--circle.  We
conclude that we did not detect the optical counterpart of
RX~J170930.2--263927 with an upper limit of m$_V=20.5$.

\subsection{Spectral analysis}
We fitted the extracted spectra using {\sc XSPEC}
(\pcite{1996adass...5...17A}) version 11.2.01.  In the spectral fits
an extra multiplicative absorption component (called {\sl ACISABS})
was included to account for absorption caused by contamination of the
ACIS optical blocking filters \footnote{see
  http://asc.harvard.edu/cal/Acis/Cal\_prods/qeDeg/}. This {\sl
  ACISABS} model is only accurate to $\sim$10 per cent. Hence, we
included a 10 per cent systematic uncertainty to the channels below 1
keV (channels 1-62). We note that, below, with the word ``absorption''
we denote the composite of interstellar absorption and this extra
absorption component. However, the N$_H$ we quote is only interstellar
absorption, unless otherwise mentioned.

\begin{figure*}
  \includegraphics[width=12cm,angle=-90]{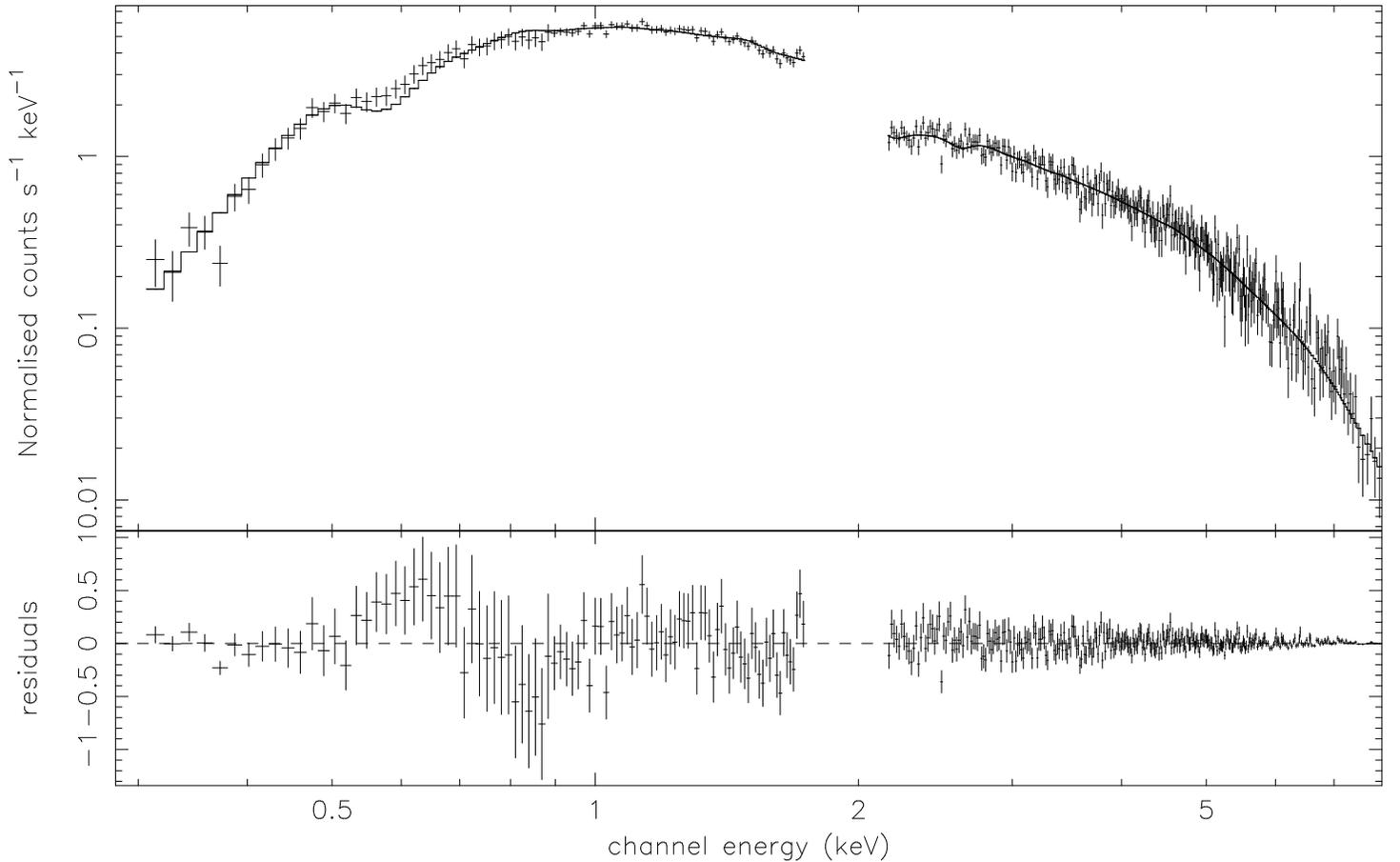}
\caption{{\em Upper panel:} {\it Chandra} X--ray spectrum (0.35--8 keV; 
  observation 2) of the neutron star SXT RX~J170930.2--263927. The
  best--fit model consisting of a linear combination of two
  blackbodies and a power law component modified by the combined
  effects of interstellar absorption and absorption due to
  contamination of the optical blocking filters of the ACIS instrument
  is overplotted. This contamination forced us to add an uncertainty
  of 10 per cent to the data for channels below 1 keV. {\em Lower
    panel:} Data minus model residuals in units of counts s$^{-1}$
  keV$^{-1}$. The excess emission near $\sim$0.6 keV is clearly
  visible. }
\label{spectra}
\end{figure*}

The fit results for the first two observations were statistically
unacceptable for two component models such as an absorbed blackbody
plus a power law model or an absorbed blackbody plus thermal
Bremsstrahlung model.  The largest residuals appear in the soft part
of the spectrum. Adding another blackbody component to the
fit--function improved the fit significantly, e.g. for the first
observation a fit using a model consisting of two absorbed blackbodies
plus a power law gives $\chi^2_{red}=$ 1.05 for 461 degrees of freedom
(d.o.f.), whereas a fit using a single absorbed blackbody and power
law model gives $\chi^2_{red}=$ 1.47 for 463 d.o.f.  Therefore, we
studied in detail a model consisting of two blackbodies and a power
law component, all reduced by absorption. Although the fits of the
models were formally acceptable (see the reduced $\chi^2$ and the
d.o.f. given in Table~2), systematic residuals near $\sim$0.6 keV were
apparent in the first two observations (e.g. see
Figure~\ref{spectra}). Such residuals have been seen before in the
X--ray spectra of several LMXBs (see \pcite{2001ApJ...560L..59J} and
references therein for figures showing residuals at $\sim$0.6 keV).
These residuals have been fitted with a Gaussian emission line
(\pcite{1994ApJ...422..791C}; \pcite{1995ApJ...449L..41A}) and, more
recently, by varying the O/Ne abundance of the circumstellar material
(\pcite{2001ApJ...560L..59J}; \pcite{2002APS..APRN17095M}). We used a
Gaussian line to account for this excess emission near 0.6 keV.  The
best--fit parameters of the model for each observation, including a 10
per cent uncertainty below 1 keV, are given in Table~2. The quoted
errors on the best--fit parameters correspond to 68\% confidence
(1~$\sigma$).

In the first two observations (where the signal--to--noise is the
highest) the inclusion of a Gaussian line centered near $\sim$0.6 keV
results in a better fit. Using an F--test to assess the significance
of the reduction in the $\chi^2$ for the fit with and without a
Gaussian gives an F statistic value of 4.7 for observation 1 and 10.8
for observation 2. The probability that the F values decrease by that
amount just by chance when the numbers of d.o.f.  decrease by two or
three for the Gaussian is $\approxlt1\times10^{-3} $ and
$\approxlt8\times10^{-7} $ for observation 1 and 2, respectively (in
the fit of the first observation the line was weak and we fixed the
FWHM to 50 eV; see Table~\ref{spec}).

\begin{sidewaystable*}

\setcounter{table}{2}
{\scriptsize
\label{spec}
\begin{tabular}{lccccccccccc}
{\normalsize {\bf Table 2.}} &{\normalsize Spectral fits$^{a,b}$ } & & & & & & & & & &  \\
& & & & & & & & & & & \\

\hline
Obs. &  PHABS & BB & & BB & & PL & & & Gaussian &       & \\
&         $N_H^h$ & kT & $^c$R$^2$/D$^2_{10}$ & kT & $^c$R$^2$/D$^2_{10}$ &    & & 
E$_l$    & FWHM  & A$_g$&  \\
&($\times10^{22}$ cm$^{-2}$) & (keV) & (km) & (keV) & (km) & $\Gamma$
& A$^d_{pl}$ & (keV) & (eV) &  & $\chi^2/$d.o.f. \\
\hline
\hline
1 &  0.44$\pm0.02$ & 0.60$\pm$0.01 & 29$\pm$3 & 0.132$\pm$0.004& 
(2.9$_{-1.0}^{+2.1}$)$\times10^{4}$& 1.31$\pm$0.08 & (9.4$\pm$0.8) & ... & ... & ... & 
1.05/461 \\
2 & 0.44$^e$ & 0.52$\pm$0.02 & 25$\pm$6 & 0.115$\pm$0.002 & $(8.6\pm1.0)\times10^4$ & 
1.7$\pm0.1$ & 14$^{+0.7}_{1.7}$ & ... & ... & ... &  1.03/404\\
3 & 0.44$^e$ & 0.88$\pm$0.09 & 0.20$^{+0.10}_{-0.07}$ & 0.22$\pm$0.01 & 139$\pm30$ & 
... & ... & ... & ... & ... & 1.03/103 \\
4 & 0.44$^e$ & 0.38$\pm$0.05 & 1.1$^{+1.0}_{-0.6}$ & 0.12$\pm$0.01 & 
800$^{+800}_{-400}$ & ... & ... & ... & ... & ... & 1.07/19 \\
5$^g$ & 0.44$^e$ & 0.22$\pm$0.03 & 14$^{+8}_{-6}$ & ... & ... & ... & ... & ... & ... & 
... & $^g$ \\
1$^f$ & 0.44$^e$ & 0.60$\pm$0.01 & 28$\pm$3 & 0.139$\pm$0.004 & 
(2.1$\pm0.3$)$\times10^4$ & 1.28$^{+0.03}_{-0.11}$ & 8.8$\pm$0.7 & 0.60$\pm$0.02 & 
50$^e$ & (8$\pm$2)$\times10^{-3}$ & 1.02/457 \\
2$^f$ & 0.44$^e$ & 0.54$\pm$0.01 & 31$\pm$5 & 0.136$\pm$0.006 & 
(2.8$\pm$0.8)$\times10^4$ & 1.5$\pm$0.2 & 9.3$\pm$2.0 & 0.57$\pm$0.02 & 
83$_{-13}^{+19}$ & (3.0$\pm$0.5)$\times10^{-2}$ & 0.96/401 \\

\hline

\end{tabular}
}

\bigskip

{\footnotesize $^a$ All quoted errors are at the 68\% confidence level
  (1~$\sigma$ single parameter). }\newline
{\footnotesize $^b$ PHABS = interstellar absorption, PL = power law, BB= 
blackbody}\newline
{\footnotesize $^c$ Blackbody normalisation in units of the radius of the emitter 
squared divided by the distance in units of 10 kpc squared. }\newline
{\footnotesize $^d$ Power Law normalisation at 1 keV in units of
  10$^{-3}$ photons keV$^{-1}$ cm$^{-2}$ s$^{-1}$.}\newline
{\footnotesize $^e$ Parameter was fixed at this value during the fit. }\newline
{\footnotesize $^f$ Including a Gaussian line in the fit to account for the excess 
near 0.6 keV }\newline
{\footnotesize $^g$ C--statistics used}\newline
{\footnotesize $^h$ The value of the interstellar absorption is determined taking the local absorption due to the {\it Chandra} optical blocking filters into account.  }\newline

\end{sidewaystable*}

Due to the low number of counts per bin during the last observation,
we used the C--statistic to estimate the goodness of fit and the error
bars on the fit parameters. The data were fit both with a blackbody
(see Table 2) and with an absorbed neutron star hydrogen atmosphere
model (\pcite{1991MNRAS.253..193P}; \pcite{1996A&A...315..141Z}). The
mass and the radius of the neutron star were held fixed in the neutron
star hydrogen atmosphere fit at 1.4 M$_\odot$ and 10 km, respectively.
Leaving the radius of the neutron star as a free parameter rendered it
unconstrained (the hard bound of 20 km of the neutron star hydrogen
atmosphere model was reached). The best--fit neutron star effective
temperature was (1.3$\pm$0.2)$\times10^6$ K.

We defined a hard colour by taking the logarithm of the ratio between
the count rates in the 5 to 8 keV and that in the 2 to 5 keV band (see
Table~\ref{fluxunabs}).

\subsection{Temporal analysis}

\begin{figure*}
  \includegraphics[width=10cm]{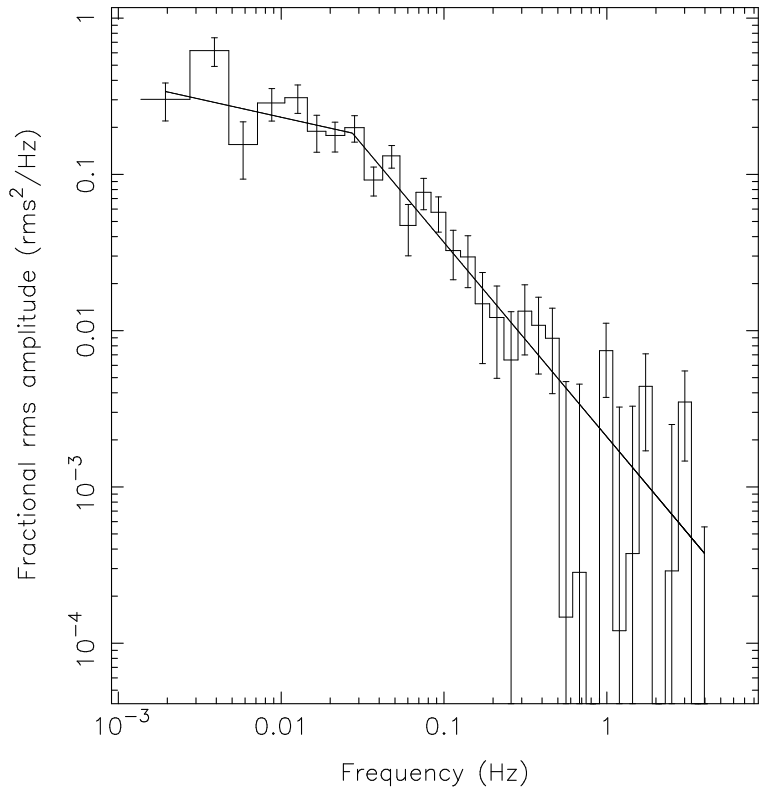}
\caption{The Poisson--noise subtracted average power density spectrum 
  (0.01--10 keV; 1/512--4 Hz) of the first {\it Chandra} observation
  (ID 3462). The drawn line represents the best--fit broken power law
  to the data. }
\label{pds}
\end{figure*}

We created power density spectra of data stretches of 512~s in length
with a Nyquist frequency of 4 Hz using the CC--mode data and with a
Nyquist frequency of 0.2 Hz using the windowed ACIS--TE mode data. All
power spectra were added and averaged for each observation. No
dead--time corrections have been applied. The average Poisson noise
level was determined by averaging the power in the 1.4--3.9 Hz region.
This average white noise level was subtracted.  We fitted the power
density spectra with a broken power law. For the third, forth, and
fifth observation only an upper limit on the fractional rms amplitude
of the broken power law component of 22, 42, and 94 per cent,
respectively, could be determined (fixing the break frequency at 0.01
Hz, the power law index below/above the break at 0/1). In the average
power spectrum of the first observation (see Figure~\ref{pds}) the
power law index below the break frequency of
(2.7$\pm$0.4)$\times10^{-2}$ Hz was 0.2$\pm$0.1, while that above the
break was 1.2$\pm$0.1.  The fractional rms amplitude (integrated from
1/1000--10 Hz) was 15$\pm$1 per cent.  The parameters of the average
power spectrum obtained during the second observations were
0.1$\pm$0.2 for the power law index below the break, 1.7$\pm$0.3 for
that above the break, (2.3$\pm$0.5)$\times10^{-2}$ Hz for the break
frequency, while the fractional rms amplitude was 12$\pm$1 per cent.
Errors on the fit--parameters were determined using $\Delta\chi^2=1.0$
(1 $\sigma$ single parameter), whereas upper limits on the broken
power law component were determined using $\Delta\chi^2=2.71$ (95 per
cent confidence).

\subsection{Burst properties}

\begin{figure*}
  \includegraphics[width=12cm]{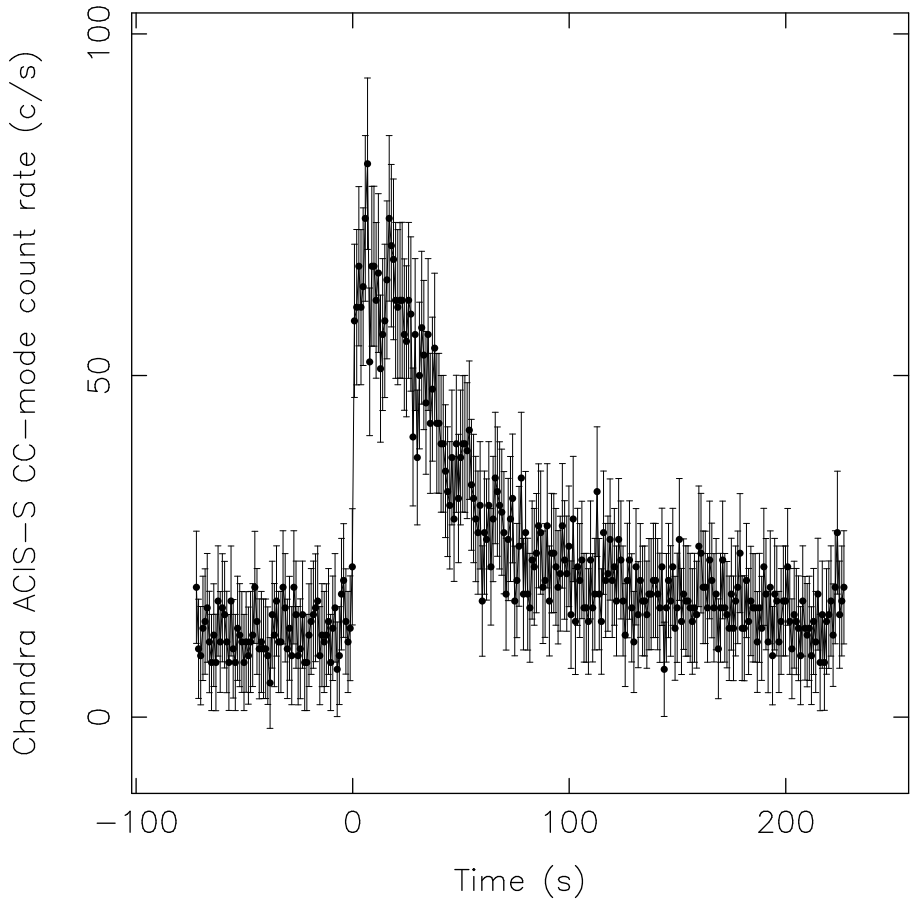}
\caption{Lightcurve of the burst in observation 1. Time zero corresponds to burst onset MJD 52345.73059 (TT).  From X--ray spectral analysis it was found that it is likely that the observation suffered from severe pile--up. The burst e--folding time is 38$\pm$3 s.}
\label{burst}
\end{figure*}

At the end of the first observation an X--ray burst occurred (see
Figure~\ref{burst}).  Inspecting the lightcurve we noted a slowdown in
the increase and even a decrease in count rate during the first stages
of the burst. To investigate whether this is due to radius expansion
or due to effects of pile--up we analysed the spectra of the burst by
combining data 0--5~s, 5--10~s, 10--15~s, 15--20~s, 20--40~s,
40--60~s, 60--80~s, and 80--160~s after burst--onset. This also allows
us to investigate whether there are signs of cooling estabilishing the
nature of the X--ray burst as an type~I X--ray burst. We subtracted the
average persistent emission as observed during the first part of the
observation from the burst data.  We found that during the first
$\sim$15 seconds of the burst the blackbody temperature was
unconstrained (best--fit values were as high as 170--190 keV). We
conclude that due to the high count rate during the first 15 to 20
seconds of the burst even in CC--mode the data suffered from severe
pile--up. Therefore, the dip in the count rate during the rising
phases of the burst could well be due to pile--up.  We therefore
excluded the first 20 seconds of the burst from further analysis.

\begin{figure*}
  \includegraphics[width=9cm,angle=-90]{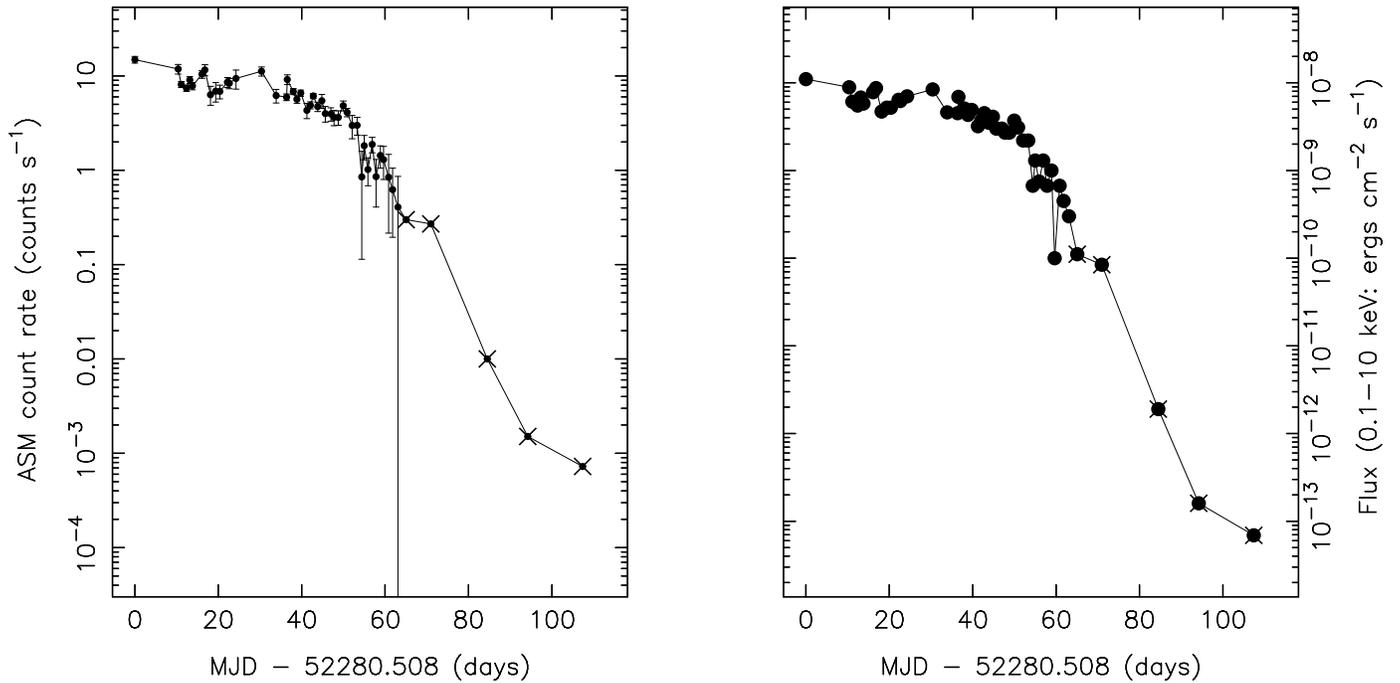}
\caption{ASM count rate ({\it left panel}) and flux ({\it right
  panel}) evolution of the 2002 outburst of RX~J170930.2--263927. The
  five crosses in each of the panels are derived from the {\it
  Chandra} measurements.  The decay profile is two--fold. Note that
  {\it (i)} given the uncertainties involved in converting flux to ASM
  counts and vice versa (see text) no error bars are given for the
  {\it right panel} nor for the five {\it Chandra} points in the {\it
  left panel} {\it (ii)} the fluxes are not corrected for the
  interstellar absorption}.
\label{decay}
\end{figure*}

The best--fit blackbody spectral parameters for the remainder of the
burst are given in Table~\ref{bfit}. Clearly, the blackbody
temperature decreased during the decay of the burst. This establishes
that the burst is a type~I X--ray burst, although 20--40 seconds after
burst onset the count rate was still high enough to cause some
pile--up (the mean count rate in that data segment was 38.1$\pm$1.6
counts per second, which results in a pile--up fraction of a few per
cent for the CC--mode). We fitted an exponential decay model to the
1~second lightcurve profile; we found an e--folding time of 38$\pm$3
seconds (excluding burst data before t$\sim$20~s).

\begin{table*}
\caption{Best fit parameters of the blackbody fits to the type~I 
X--ray burst in RX~J170930.2--263927$^a$}
\label{bfit}

\begin{tabular}{lccc}
\hline
Data segment & kT & $^c$R$^2$/D$^2_{10}$ & 0.1--10 keV flux \\
Time (s)$^b$ & (keV) & (km) & (erg cm$^{-2}$ s$^{-1}$)    \\
\hline
\hline
20--40 & 2.1$\pm$0.1 & 5.6$\pm$1.0 & 8.1$\times10^{-10}$ \\
40--60 & 1.3$\pm$0.1 & 9.5$\pm$2.0 & 2.8$\times10^{-10}$ \\
60--80 & 1.3$\pm$0.1 & 4.9$\pm$1.0 & 1.3$\times10^{-10}$ \\
80--160 &1.15$\pm0.10$ & 3.1$\pm$0.8 & 5.8$\times10^{-11}$ \\

\hline
\end{tabular}

\bigskip

{\footnotesize $^a$ All quoted errors are at the 68\% confidence level
  (1~$\sigma$ single parameter). }\newline
{\footnotesize $^b$ Times are given with respect to burst onset.}\newline
{\footnotesize $^c$ Blackbody normalisation in units of the radius of the emitter 
squared divided by the distance in units of 10 kpc squared. }\newline
\end{table*}

\subsection{RXTE/ASM observations}

To determine the duration and decay profile of the outburst we
obtained data from RXTE's All Sky Monitor (ASM; \pcite{brrosw1993};
\pcite{lebrcu1996}) for the 2002 outburst of RX~J170930.2--263927.
Using PIMMS and a power law model with index 2 as input we converted
the ASM count rate to a flux in the 0.1--10 keV band. From previous
observations of other LMXBs and our own {\it Chandra} observations of
this source it is known that the source spectrum changes as the
outburst progesses and that such a single--component spectrum can
serve only as a first approximation of the spectral shape.  Hence, the
flux determinations from the ASM count rates are provided as reference
only (Figure~\ref{decay}, {\it right panel}).

From the best--fit spectra we derived the source flux in the 0.1--10
keV energy band for each of the {\it Chandra} observations; these are
the five rightmost points in each of the panels of Figure~\ref{decay}.
In Table~\ref{fluxunabs} we provide the {\it unabsorbed} fluxes for
the five {\it Chandra} observations.  Using PIMMS we converted these
{\it Chandra} fluxes assuming a blackbody model temperature of 0.3 keV
into ASM count rates (Figure~\ref{decay}, {\it left panel}). The
source outburst decay is two--fold. Until MJD $\sim$52331 the decay is
slow, but after that the rate of decay increases by a factor $\sim10$
(a fit of a broken power law to the ASM count rate data gives a power
law index of $(-7.6\pm1.4)\times10^{-3}$ before $t=50\pm1$ days and
$(-7.4\pm0.2)\times10^{-2}$ after that).

\begin{table*}

\caption{Unabsorbed flux measurements of the best--fit spectral model
and the hard colour (see text) of the five {\it Chandra} observations. }

\label{fluxunabs}

\begin{tabular}{lccc}
\hline
Observation & flux (0.1--10 keV) & luminosity & hard colour \\
            &  (erg cm$^{-2}$ s$^{-1}$) & $(\frac{{\rm d}}{10{\rm kpc}})^2$ erg s$^{-1}$ & \\
\hline
\hline
1 &  1.8$\times10^{-10}$ & 2.0$\times10^{36}$& -0.56\\
2 &  1.9$\times10^{-10}$ & 2.1$\times10^{36}$& -0.60\\
3 &  3.9$\times10^{-12}$ & 4.4$\times10^{34}$& -0.85\\
4 &  9.9$\times10^{-13}$ & 1.1$\times10^{34}$& -2.59\\
5 &  2.5$\times10^{-13}$ & 2.8$\times10^{33}$& -4.57\\
\hline
\end{tabular}
\end{table*}

\subsection{Archival ROSAT spectra}

The ROSAT satellite observed and detected RX~J170930.2--263927 twice.
The source was first detected in the ROSAT All Sky Survey observations
on August 21, 1990 (\pcite{1999A&A...349..389V}).  We analysed these
archival ROSAT data using the {\sc ftool Xselect} (LHEASOFT version
5.2). The 0.2--2.4 keV ROSAT spectrum was fitted with an absorbed
blackbody.  Fixing the amount of absorption to the value we found
using the {\it Chandra} observations (N$_H=0.44\times10^{22}$
cm$^{-2}$) a blackbody temperature of 0.22$\pm$0.01 keV and radius of
(2$\pm$0.5)$\times10^2$ km were found.  The unabsorbed source flux (0.1--10
keV) was 4.8$\times10^{-12}$ erg cm$^{-2}$ s$^{-1}$.

The source was also in the field of view of a $\sim$900~s pointed PSPC
observation of the source V2051~Oph obtained on September 22, 1992
(see also \pcite{2001A&A...368..137V}). The 0.2--2.4 keV source
spectrum was well fit by an absorbed blackbody of a temperature of
0.22$\pm$0.02 keV and a radius of (2.9$\pm$1.0)$\times10^3$ km (again
the N$_H$ was fixed at $0.44\times10^{22}$ cm$^{-2}$). The unabsorbed
source flux (0.1--10 keV) was 7.3$\times10^{-11}$ erg cm$^{-2}$
s$^{-1}$.

\section{Discussion}

We observed the soft X--ray transient (SXT) RX~J170930.2--263927 five
times with the {\it Chandra} satellite in March-- April 2002 after an
X--ray outburst. The spectrum of the source was well fit by a model
consisting of two blackbodies and a power law, all absorbed by
interstellar absorption (taking into account excess absorption due to
contamination of the ACIS optical blocking filters.) The best--fit
blackbody temperatures and radii (see Table 2) imply that the
accretion disk contribution is soft (blackbody temperature of 0.1--0.2
keV) and that there is a contribution from a hotter (blackbody
temperature of $\sim$0.6 keV), smaller site, possibly the neutron star
or the neutron star boundary layer. In the accreting millisecond
pulsar XTE~J0929--314 an 0.6 keV blackbody component has been found as
well (\pcite{2002ApJsubm}), whereas in XTE~J1751--305 the blackbody
temperature is $\sim$1 keV (\pcite{2002ApJMiller1751}). With a power
law index of 1.1--1.5 the power law index is somewhat harder than
typical for LMXBs; they have indices's of $\sim$2
(\pcite{whnapa1995}; \pcite{1997ApJS..109..177C}). 

The spectral fits of the first two {\it Chandra} observations showed
evidence for excess emission at $\sim$0.6 keV. Such an excess has been
found in other LMXBs as well and has been identified with a blend of
emission lines from O~VII--O~VIII and/or Fe XVII--Fe~XIX
(\pcite{1994ApJ...422..791C}). Recently, \scite{2001ApJ...560L..59J}
showed that this excess can also be explained by an enhanced Ne/O ratio
with respect to solar. Since two of the sources which show such an
excess have an ultra short orbital period and the other sources have
low absolute visual magnitudes, it was proposed that this enhanced
Ne/O ratio is due to the fact that the compact objects are accreting
matter of a neon--rich degenerate dwarf in ultracompact systems. In
RX~J170930.2--263927 the excess can be well--fit by a Gaussian line
with an equivalent width of $\sim$50--80 eV centred at $\sim$0.59 keV.

Could the extra emission component in the spectrum be due to the
enhanced absorption due to the contamination of the optical blocking
filters? Tests performed by the {\it Chandra} X--ray Center show that
the model describing the excess absorption as a function of time is
better than 10\% except near the C K--edge but this edge falls outside
the spectral range we considered. We included this 10 per cent as a
systematic uncertainty in our fits for channels below 1 keV; still we
found excess emission near 0.6 keV. We conclude that it is unlikely
that this excess emission is due to the imperfect correction for the
contamination of the optical blocking filters. Hence, if the 0.6
keV excess is a unique feature of ultracompact X--ray binaries
RX~J170930.2--263927 has a short orbital period as well.

So far, two outbursts of RX~J170930.2--263927 have been observed with
RXTE's ASM since the launch of RXTE (December 1995). Both outburst
profiles show a steepening of the decay approximately 50 days after
the start of the outburst (see for a figure of the first outburst
profile \pcite{1998ApJ...508L.163C}, and Figure~\ref{decay} for the
profile of this outburst). Such a steepening in the decay of the
outburst has been observed in other SXTs as well; this steepening has
been interpreted as evidence for the onset of the propeller mechanism
(\pcite{1975A&A....39..185I}; Aql~X--1, \pcite{1998ApJ...499L..65C};
SAX~J1808.4--3658, \pcite{1998A&A...338L..83G}). However, since type~I
X--ray bursts are observed after the alleged onset of the propeller
mechanism (\pcite{1998ApJ...508L.163C}; this work) this means that
mass accretion must be ongoing after the steepening of the decay.
Furthermore, \scite{1999ApJ...521..332P} found that in the accreting
millisecond X--ray pulsar SAX~J1808.4--3658 pulsation were found after
the supposed onset of the propeller effect. Together, these findings
pose a serious problem for the interpretation of the steepening of
the decay as being due to the onset of the propeller mechanism.
No pulsations have been reported for RX~J170930.2--263927 (we
determined a 95 per cent upper limit to pulsations in the frequency
range 100--1000 Hz of 4.4 per cent fractional rms amplitude using the
archival RXTE/PCA observation [2--60 keV; ID 20208-01-02-00] obtained
during the previous outburst).

From the spectral fits and the hardness ratio (see
Table~\ref{fluxunabs}) it was found that the source spectrum softened
considerably between the third and fourth observation. Typically the
spectra of SXTs become soft when $L<10^{34}$ erg s$^{-1}$
(\pcite{1996ARA&A..34..607T}). It is unclear what causes this
softening, but it has been shown that the spectra of accreting neutron
stars become soft for low mass accretion rates
(\pcite{1995ApJ...439..849Z}).

The best--fit interstellar absorption (0.44$\times10^{22}$cm$^{-2}$)
we found is somewhat higher than the interstellar absorption expected
on the basis of the source location (\pcite{1990ARA&A..28..215D}). A
more accurate calculation of the interstellar absorption in the
direction of RX~J170930.2--263927 based on the work of
\scite{scfida1998} finds N$_H=$0.34$\times10^{22}$cm$^{-2}$. Possibly,
some of the absorption is local to the system. Given the high Galactic
latitude of the source this provides a lower limit to the source
distance of 2.5 kpc.

We detected a burst from RX~J170930.2--263927. Due to effects of
pile--up, the first 20~s of the burst could not be used for spectral
analysis.  Analysis of the data obtained during the later stages of
the burst showed evidence of spectral cooling, making it likely that
this is a type~I X--ray burst, establishing the nature of the compact
object of this transient as a neutron star (confirming the findings
of \pcite{1998ApJ...508L.163C}). The burst e--folding time of 38$\pm$3
seconds makes it a typical mixed He/H burst; this is not contradicting
the possibility that RX~J170930.2--263927 is accreting from a hydrogen
deficit donor star as \scite{1992ApJ...384..143B} showed that nearly
all metals will be destroyed in the atmosphere of the neutron star due
to spallation processes. \scite{1998ApJ...508L.163C} derived an upper
limit to the distance of RX~J170930.2--263927 of 10$\pm$1 kpc by
assuming that the peak burst flux is less than the Eddington
luminosity for a 1.4 M$_\odot$ neutron star (L$_{{\rm Edd}} \sim
2\times10^{38}$ erg s$^{-1}$). Approximately the same upper limit is
derived assuming that the peak {\it outburst} flux (1--2$\times10^{-8}$
\,${\rm erg\,cm^{-2} s^{-1}}$) is less then the Eddington limit for a 1.4
M$_\odot$ neutron star. If the distance to RX~J170930.2--263927 is
less then 4--5 kpc the source outburst luminosity is low and the
source outburst luminosity fits in with that of the proposed class of
faint SXTs (\pcite{1999ApL&C..38..297H}; \pcite{2001ESASP.459..463I}).

We analysed data of RX~J170930.2--263927 obtained by the ROSAT
satellite. ROSAT observed and detected the source twice; once in
August 1991 at a 0.1--10 keV unabsorbed flux level of
4.8$\times10^{-12}$ erg cm$^{-2}$ s $^{-1}$ (L$=5.4\times10^{34}$
$(\frac{{\rm d}}{10{\rm kpc}})^2$ erg s$^{-1}$), and once in September
1992 at an unabsorbed flux level (0.1--10 keV) of 7.3$\times10^{-11}$
erg cm$^{-2}$ s $^{-1}$ (L$=8.3\times10^{35}$ $(\frac{{\rm d}}{10{\rm
    kpc}})^2$ erg s$^{-1}$).  Since the possiblity that ROSAT observed
the source during an outburst is low given the fact that the ASM only
observed two outbursts in over 6 years, we conclude that
RX~J170930.2--263927 must either have had a much shorter outburst
interval time when the ROSAT observations were made, or that the
source is active for a long time at a low luminosity level, similar to
the neutron star SXTs SAX~J1808.4--3658 (\pcite{2001ApJ...560..892W}),
4U~1608--52 (\pcite{2002ApJ...568..901W}), Aql~X--1
(\pcite{2002ApJ...568..901W}), SAX~J1747.0--2853
(\pcite{2002ApJ...579..422W}), and EXO~0748--676.  Variability of
ongoing low--level accretion in these transient systems could provide
an alternative explanation for some of the {\it faint} SXT sources
(\pcite{1999ApL&C..38..297H}). As was explained by
\scite{2002A&A...392..931C} low level accretion in transient sources
and variability therein could also explain the properties of several
of the faint burst sources detected with the {\it Beppo}SAX satellite.
Variability up to a factor of 10 has been observed during ongoing
low--level accretion in 4U~1608--52 (\pcite{2002ApJ...568..901W})
whereas much larger variations (up to a factor 1000 in 5 hours) have
been observed in SAX~J1808.4--3658 (\pcite{2001ApJ...560..892W}).
Black hole candidate SXTs have also shown activity after the decay of
an outburst (e.g. 4U~1630--47, \pcite{1997MNRAS.291...81K}).

It is unclear whether RX~J170930.2--263927 was already in quiescence
during our last {\it Chandra} observation. The source luminosity
during the last observation was between
$1.8\times10^{32}$--$2.8\times10^{33}$ erg s$^{-1}$ for a distance of
2.5--10 kpc, respectively. This luminosity range is consistent with
the luminosity expected from cooling of a neutron star core
(\pcite{1998ApJ...504L..95B}), although for a distance of 2.5 kpc the
luminosity is rather low. The ratio between the outburst flux and
quiescence flux is $\approxgt10^5$ (see Fig.~\ref{decay}); this is
similar to that of Cen~X--4. The outburst recurence time is $\sim$5
years and the outburst duration is $\sim$100 days (as determined from
the RXTE/ASM records). Possibly, the time averaged mass accretion rate
is an order of magnitude (or more) lower than 10$^{-11}$ M$_\odot$
yr$^{-1}$, leaving the neutron star core relatively cold. The
quiescent core luminosity after the neutron star crust has cooled
should be very low (10$^{30-31}$ erg s$^{-1}$, for a time averaged
mass accretion rate of 10$^{-13}-10^{-12} M_\odot$ year$^{-1}$,
respectively). Such a low time averaged mass accretion rate is not
unfeasible for ultracompact systems (e.g. see
\pcite{2002A&A...388..546Y}). However, the fluence of the March--April
2002 outburst of RX~J170930.2--263927 is $\sim0.02$ erg cm$^{-2}$. For
a recurrence time of five years this leads to a time averaged flux of
1.3$\times10^{-10}$ erg cm$^{-2}$ s$^{-1}$ and
$\langle$\.{M}$\rangle\sim1.2\times10^{-10}$ M$_\odot$ yr$^{-1}$
$(\frac{{\rm d}}{10 {\rm kpc}})^2$, making it unlikely that the
neutron star core is very cold as the result of a very low
time--averaged mass accretion rate. Alternatively, enhanced neutrino
cooling of the core (\pcite{2001ApJ...548L.175C}) may take place; this
could occur for massive neutron stars (1.7--1.8 M$_\odot$). The low
quiescent luminosity of the accreting millisecond X--ray pulsar
SAX~J1808.4--3658 can also be explained by enhanced core cooling
(\pcite{2002ApJ...575L..15C}).

From the {\it Chandra} ACIS--TE observation we derived an accurate
position for RX~J170930.2--263927. We note that this position is
$\sim$9--10 arcminutes away from the centre of the metal--poor Globlar
Cluster NGC~6293 at a distance of 8.5 kpc (although the uncertainties
in the cluster distance are considerable;
\pcite{1991AJ....101.2097J}). This Globular Cluster has a tidal radius
of 14.2 arcminutes (\pcite{1996AJ....112.1487H}). It is not unfeasible
that the source was slung out of the Globular Cluster by interactions
with other more massive components in the Globular Cluster. If the
association is real and not a chance alignment, RX~J170930.2--263927
has a projected distance of $\sim25$ pc from the centre of the
cluster. If the source is unassociated with the Globular Cluster it
has travelled from the plane of the Galaxy.  E.g. for a distance of
8.5 kpc RX~J170930.2--263927 is 1.2 kpc above the Galactic plane;
assuming RX~J170930.2--263927 was born in the Galactic plane and it is
at the highest point in its Galactic orbit, the initial space velocity
must at least have been 120 km s$^{-1}$ (we used the potential for the
Galaxy used by \pcite{lorri94} in the implementation of
\pcite{1997A&A...322..477H}).

We did not detect the optical counterpart with a five sigma upper
limit of m$_V \sim$20.5. The upper limit on the absolute magnitude of
the counterpart to RX~J170930.2--263927 is $\sim$3.5 for an optical
visual extinction of 2.4 magnitudes (taking N$_H=0.44\times10^{22}$
cm$^{-2}$, the relation between N$_H$ and A$_V$ of
\pcite{1995A&A...293..889P}, and a distance of 8.5 kpc [if the source
is closer than that the absolute optical magnitude will be larger]).
This absolute optical magnitude is large for an LMXB (cf.
\pcite{vamc1994}). Although, since our optical observations were
performed when RX~J170930.2--263927 was most likely in quiescence it
is difficult to compare this absolute optical magnitude with the work
of \scite{vamc1994} as those authors considered actively accreting
sources. During periods of accretion SXTs are known to be much
brighter in optical than in quiescence (in case of the black hole
candidate A~0620--00 the optical magnitude decreased by $\sim$7
magnitudes, see \pcite{1995xrb..book...58V}). Since it is unknown by
how much RX~J170930.2--263927 brightened in outburst it is unclear
what the absolute optical magnitude was. However, from the upper limit
on the absolute visual magnitude we derive that the spectral type of
the companion must be later than F5--G0, making RX~J170930.2--263927 a
low--mass X--ray binary, unless the distance to the source is much
larger than 8.5 kpc.

\section*{Acknowledgments} 
\noindent 
We would like to thank the {\it Chandra} Director, Harvey Tanenbaum,
for approving these DDT observations and Yousaf Butt for help with
selecting the best observation modes. PGJ is supported by EC Marie
Curie Fellowship HPMF--CT--2001--01308. MK is supported in part by a
Netherlands Organization for Scientific Research (NWO) grant. This
research made use of results provided by the ASM/RXTE teams at MIT and
at the RXTE SOF and GOF at NASA's GSFC. PGJ would like to thank Jean
Swank for communicating RXTE/PCA flux measurements which helped
trigger the {\it Chandra} observations. We would like to thank the
referee for his/her comments which helped improve the manuscript.

\end{document}